\documentclass[journal]{IEEEtran}
\usepackage{cite}
\usepackage[rightcaption]{sidecap}

\ifCLASSINFOpdf
   
   \usepackage[pdftex]{graphicx}
   \usepackage[skip=2pt]{caption}
   \usepackage{multirow}
   \usepackage{tabularew}
   \usepackage[flushleft]{threeparttable}
   
   \graphicspath{{./pdf/}{./jpeg/}{./eps/}}
  \DeclareGraphicsExtensions{ ,.jpeg,.png}
\else
   \usepackage[dvips]{graphicx}
   \graphicspath{{./eps/}}
   \DeclareGraphicsExtensions{.eps}
\fi
\usepackage{amsmath}
\usepackage{color,soul}

\usepackage{array}

\ifCLASSOPTIONcompsoc
 \usepackage[caption=false,font=normalsize,labelfont=sf,textfont=sf]{subfig}
\else
  \usepackage[caption=false,font=footnotesize]{subfig}
\fi

\usepackage{stfloats}
\usepackage{url}
  \usepackage{color}

\usepackage{url}

\begin{document}

\title{8T SRAM Cell as a Multi-bit Dot Product Engine for Beyond von-Neumann Computing}
%
%
%

\author{Akhilesh Jaiswal*, Indranil Chakraborty*, Amogh Agrawal, Kaushik~Roy,~\IEEEmembership{Fellow,~IEEE}
\\ *(Equal Contributor)}
\maketitle
\pagenumbering{gobble}

\begin{abstract}

Large scale digital computing almost exclusively relies on the von-Neumann architecture which comprises of separate units for storage and computations. The energy expensive transfer of data from the memory units to the computing cores results in the well-known von-Neumann bottleneck. Various approaches aimed towards bypassing the von-Neumann bottleneck are being extensively explored in the literature. These include \textit{in-memory} computing based on CMOS and beyond CMOS technologies, wherein by making modifications to the memory array, vector computations can be carried out as close to the memory units as possible. Interestingly, in-memory techniques based on CMOS technology are of special importance due to the ubiquitous presence of field-effect transistors and the resultant ease of large scale manufacturing and commercialization. On the other hand, perhaps the most important computation required for applications like machine-learning \textit{etc.} comprises of the dot product operation. Emerging non-volatile memristive technologies have been shown to be very efficient in computing analog dot products in an \textit{in-situ} fashion. The memristive analog computation of the dot product results in much faster operation as opposed to digital vector in-memory bit-wise Boolean computations. However, challenges with respect to large scale manufacturing coupled with the limited endurance of memristors have hindered rapid commercialization of memristive based computing solutions. In this work, we show that the standard 8 transistor (8T) digital SRAM array can be configured as an \textit{analog-like in-memory multi-bit dot product engine}. By applying appropriate analog voltages to the read-ports of the 8T SRAM array, and sensing the output current, an approximate analog-digital dot-product engine can be implemented. We present two different configurations for enabling multi-bit dot product computations in the 8T SRAM cell array, without modifying the standard bit-cell structure. We also demonstrate the robustness of the present proposal in presence of non-idealities like the effect of line-resistances and transistor threshold voltage variations. Since our proposal preserves the standard 8T-SRAM array structure, it can be used as a storage element with standard read-write instructions, and also as an \textit{on-demand} analog-like dot product accelerator.

\end{abstract}

\begin{IEEEkeywords}
In-memory computing, SRAMs, von Neumann bottleneck, convolution, dot product.
\end{IEEEkeywords}

%
\IEEEpeerreviewmaketitle

\section{Introduction}

\IEEEPARstart{S}tate-of-the-art computing platforms are widely based on the von-Neumann architecture \cite{von2012computer}. The von-Nuemann architecture is characterized by distinct spatial units for \textit{computing} and \textit{storage}. Such physically separated memory and compute units result in huge energy consumption due to frequent data transfer between the two entities. Moreover, the transfer of data through a dedicated limited-bandwidth bus limits the overall compute throughput. The resulting memory bottleneck is \textit{the major throughput concern} for hardware implementations of data intensive applications like machine learning, artificial intelligence \textit{etc}. 

\begin{figure*}[t]
\centering
\includegraphics[width=0.9\textwidth]{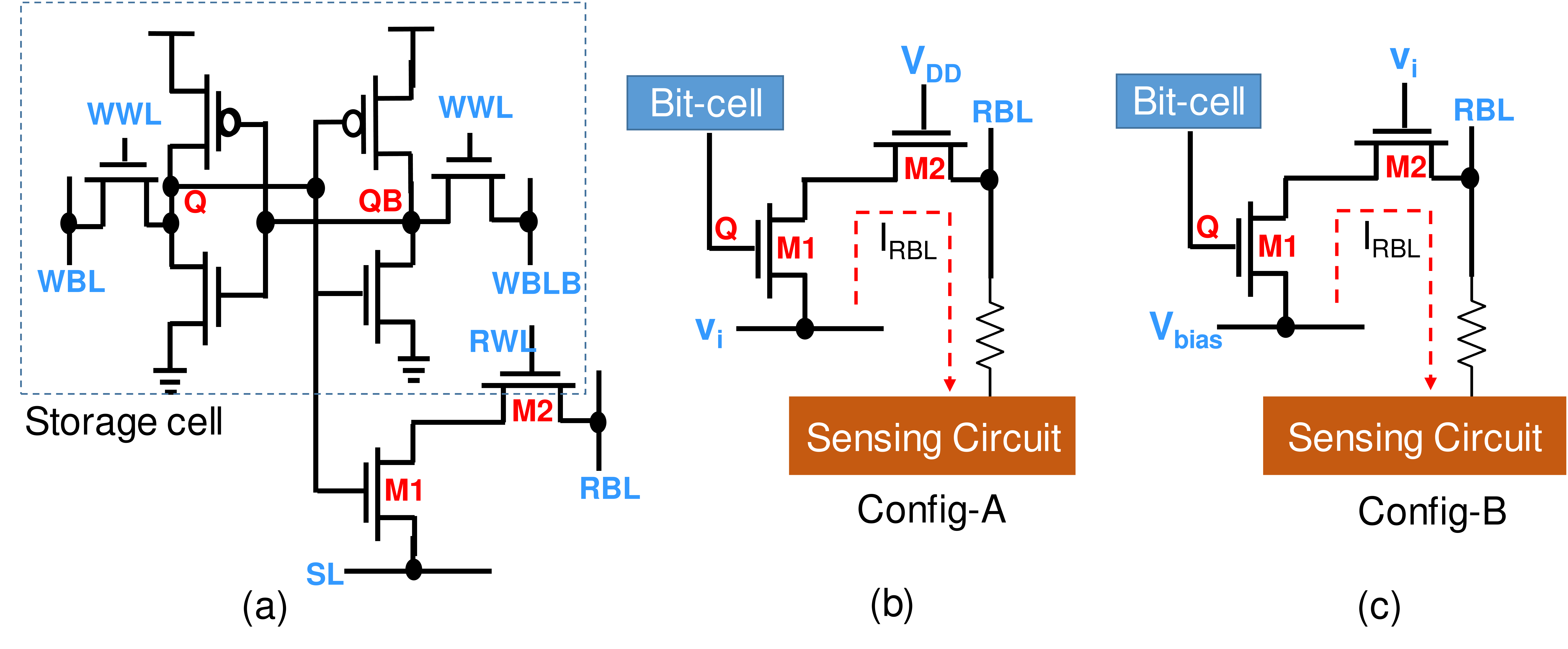}
\caption{(a) Schematic of a standard 8T-SRAM bit-cell. It consists of two decoupled ports for reading and writing respectively. (b) First proposed configuration (Config-A) for implementing the dot product engine using the 8T-SRAM bit-cell. The SL is connected to the input analog voltage $v_i$, and the RWL is turned ON. The current $I_{RBL}$ through the RBL is sensed and is proportional to the dot product $v_i \cdot g_i$, where $g_i$ is the ON/OFF conductance of the transistors $M1$ and $M2$. (c) Second proposed configuration (Config-B). The input analog voltages are applied to the RWL, while the SL is supplied with a constant voltage $V_{bias}$. The current through the RBL is sensed in the same way as in Config-A.
}
\label{fig:8t} 
\end{figure*}

A possible approach geared towards high throughput beyond von-Neumann machines is to enable distributed computing characterized by tightly intertwined storage and compute capabilities. If computing can be performed inside the memory array, rather than in a spatially separated computing core, the compute throughput can be considerably increased. As such, one could think of \textit{ubiquitous} computing on the silicon chip, wherein both the logic cores and the memory unit partake in compute operations. 
Various proposals for \textit{`in-memory'} computing with respect to emerging non-volatile technologies have been presented for both dot product computations \cite{yang2013memristive,Li_2017} as well as vector Boolean operations \cite{borghetti2010memristive}. Prototypes based on emerging technologies can be found in \cite{jo2010nanoscale,Li_2017} . 

With respect to the CMOS technology, Boolean in-memory operations have been presented in \cite{dong20170} and \cite{agrawal2017x}. In \cite{dong20170} authors have presented vector Boolean operations using 6T SRAM cells. Additionally, authors in \cite{agrawal2017x} have demonstrated that the 8 transistor (8T) SRAM cells lend themselves easily as vector compute primitives due to their decoupled read and write ports. Both the works \cite{dong20170} and \cite{agrawal2017x} are based on vector Boolean operations. However, perhaps the most frequent and compute intensive function required for numerous applications like machine learning is the \textit{dot product} operation. Memristors based on resistive-RAMs (Re-RAMs) have been reported in many works as an analog dot product compute engine \cite{borghetti2010memristive, li2018analogue}. Few works based on analog computations in SRAM cells can be found in \cite{asram1, asram2}. Both these works use 6T SRAM cells and rely on the resultant accumulated voltage on the bit-lines (BLs). Not only 6T SRAMs are prone to read-disturb failures, the failures are also a function of the voltage on the BLs. This leads to a tightly constrained design space for the proposed 6T SRAM based analog computing.
In this paper, we employ 8T cells that are much more robust as compared to the 6T cells due to isolated read port. We show that without modifying the basic bit-cell for the 8T SRAM cell, it is possible to configure the 8T cell for in-memory dot product computations. Note, in sharp contrast to the previous works on in-memory computing with the CMOS technology, we enable \textit{current based, analog-like} dot product computations using robust digital 8T bit-cells. 

The key highlights of the present work are as follows:

\begin{enumerate}

\item We show that the conventional 8T SRAM cell can be used as a primitive for analog-like dot product computations, without modifying the bit-cell circuitry. In addition, we present two different configurations for enabling dot product computation using the 8T cell.

\item Apart for the sizing of the individual transistors consisting the read port of the 8T cell, the basic bit-cell structure remains unaltered. Thereby, the 8T SRAM array can also be used for usual digital memory read and write operations. As such, the presented 8T cell array can act as a dedicated dot product engine or as an \textit{on-demand} dot product accelerator.

\item A detailed simulation analysis using 45nm predictive technology models including layout analysis and effect of non-idealities like the existence of line-resistances and variation in transistor threshold voltages has been reported highlighting the various trade-offs presented by each of the two proposed configurations.

\end{enumerate}

\section{8T-SRAM as a Dot Product Engine}

A conventional 8T bit-cell is schematically shown in Fig. \ref{fig:8t}(a). It consists of the well-known 6T-SRAM bit-cell with two additional transistors that constitute a decoupled read port. To write into the cell, the write word-line (WWL) is enabled, and write bit-lines (WBL/WBLB) are driven to $V_{DD}$ or ground depending on the bit to be stored. To read a value from the cell, the read bit-line (RBL) is pre-charged to $V_{DD}$ and the read word-line (RWL) is enabled. Note, that the source-line (SL) is connected to the ground. Depending on whether the bit-cell stores a logic `1' or `0', the RBL discharges to 0V or stays at $V_{DD}$, respectively. The resulting voltage at the RBL is read out by the sense amplifiers. Although 8T-cells incur a $\sim$30\% increase in bit-cell area compared to the 6T design, they are read-disturb free and more robust due to separate read and write path optimizations \cite{chang2005stable}. 

We now show how such 8T-SRAMs, with no modification to the basic bit-cell circuit (except for the sizing of the read transistors), can behave as a dot product engine, without affecting the stability of the bits stored in the SRAM cells. We propose two configurations - \textit{Config-A} and \textit{Config-B}, for enabling dot-product operations in the 8T-SRAMs. Config-A is shown in Fig. \ref{fig:8t}(b). The inputs $v_i$ (encoded as analog voltages) are applied to the SLs of the SRAM array, and the RWL is also enabled. The RBL is connected to a sensing circuitry, which we will describe later. Thus, there is a static current flow from the SL to the RBL, which is proportional to the input $v_i$ and the conductance of the two transistors $M1$ and $M2$. For simplicity, assume that the weights (stored in the SRAM) have a single-bit precision. If the bit-cell stores `0', the transistor $M1$ is OFF, and the output current through the RBL is close to 0. Whereas if the bit-cell stores a `1', the current is proportional to $v_i \cdot g_{ON}$, where $g_{ON}$ is the series `ON' conductance of the transistors. Assume similar inputs $v_i$ are applied on the SLs for each row of the memory array. Since the RBL is common throughout the column, the currents from all the inputs $v_i$ are summed into the RBL. Moreover, since the SL is common throughout each row, the same inputs $v_i$ are supplied to multiple columns. Thus, the final output current through RBL of each column is proportional to $I_{RBL}^j=\Sigma (v_i\cdot g_i^j)$, where $g_i^j$ is the `ON' or `OFF' conductance of the transistors, depending on whether the bit-cell in the i-th row and j-th column stores a `1' or `0', respectively. The output current vector thus resembles the vector-matrix dot product, where the vector is $v_i$ in the form of input analog voltages, and the matrix is $g_i^j$ stored as digital data in the SRAM.

\begin{figure}[t]
\centering
\includegraphics[width=0.5\textwidth]{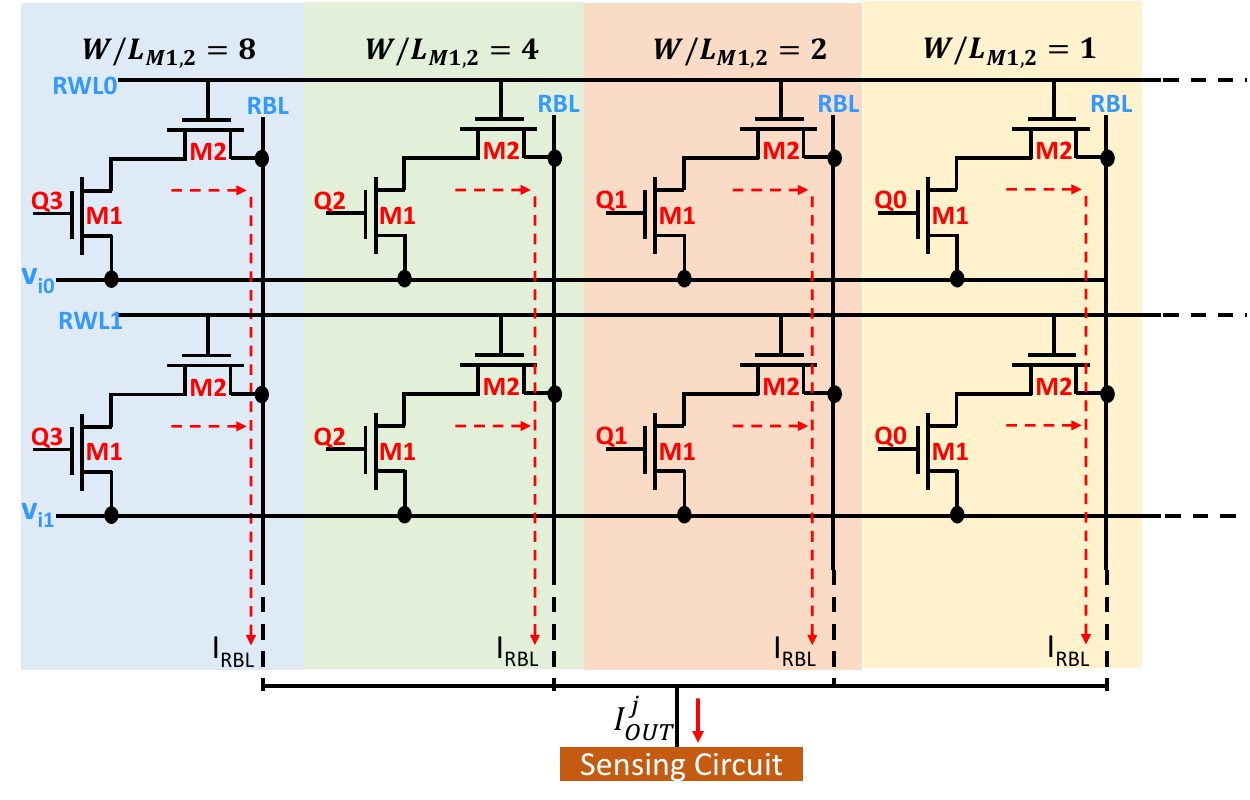}
\caption{8T-SRAM memory array for computing dot-products with 4-bit weight precision. Only the read port is shown, the 6T storage cell and the  write port are not shown. The array columns are grouped in four, and the transistors $M1$ and $M2$ are sized in the ratio $8:4:2:1$ for the four columns. The output current $I_{OUT}^j$ represents the weighted sum of the $I_{RBL}$ of the four columns, which is approximately equal to the desired dot-product.
}
\label{fig:8tarr} 
\end{figure}

Let us now consider a 4-bit precision for the weights. If the weight $W_i^j={w_3w_2w_1w_0}$, where $w_i$ are the bits corresponding to the 4-bit weight, the vector matrix dot product becomes:\\

$\Sigma (v_i\cdot W_i^j)=\Sigma [v_i\cdot (2^3w_3+2^2w_2+2^1w_1+w_0)]$\\
$=\Sigma (v_i\cdot 2^3w_3)+\Sigma (v_i\cdot2^2w_2)+\Sigma (v_i\cdot2^1w_1)+\Sigma (v_i\cdot w_0)$\\

Now, if we size the read transistors $M1$ and $M2$ of the SRAM bit-cells in column 1 through 4 in the ratio $2^3:2^2:2^1:1$, as shown in Fig. \ref{fig:8tarr}, the transistor conductances in the `ON' state would also be in the ratio $2^3:2^2:2^1:1$. Thus, summing the currents through the RBLs of the four columns yields the required dot product in accordance to the equation shown above. This sizing pattern can be repeated throughout the array. In addition, one could also use transistors having different threshold voltages to mimic the required ratio of conductances as $2^3:2^2:2^1:1$. Note that, the currents through the RBLs of the four consecutive columns are summed together, thus we obtain one analog output current value for every group of four columns. In other words, the digital 4-bit word stored in the SRAM array is multiplied by the input voltage $v_i$ and summed up by analog addition of the currents on the RBLs. This \textit{one-go} computation of vector multiplication and summation in a digital memory array would result in high throughput computations of the dot products.

It is worth mentioning, that the way input $v_i$ are multiplied by the stored weights and summed up is reminiscent of memristive dot product computations \cite{li2018analogue}. However, a concern with the presented SRAM based computation is the fact that the ON resistance of the transistors (few kilo ohms) are much lower as compared to a typical memristor ON resistance which is in the range of few tens of kilo ohms \cite{chakraborty2017technology}. As such the static current flowing through the ON transistors $M1$ and $M2$ would typically be much higher in the presented proposal. In order to reduce the static current flow, we propose scaling down the supply voltage of the SRAM cell. Note, interestingly, 8T cells are known to retain their robust operation even at highly scaled supply voltages \cite{chang20075}. In the next section we have used a $V_{DD}$ lower than the nominal $V_{DD}$ of 1V. We would now describe another way of reducing the current, although with trade-offs, as detailed below. 


Config-B is shown in Fig. \ref{fig:8t}(c). Here, the SLs are connected to a constant voltage $V_{bias}$. The input vector $v_i$ is connected to RWLs, i.e., the gate of $M2$. Similar to Config-A, the output current $I_{RBL}$ is proportional to $v_i$. We will later show from our simulations that for a certain range of input voltage values, we get a linear relationship between $I_{RBL}$ and $v_i$, which can be exploited to calculate the approximate dot product. To implement multi-bit precision, the transistor sizing is done in the same way as Config-A as represented in Fig. \ref{fig:8tarr}, so that the $I_{RBL}$ is directly proportional to the transistor conductances. Key features of the proposed Config-B are as follows.  The input voltages $v_i$ have a capacitive load, as opposed to a resistive load in Config-A. This relaxes the constraints on the input voltage generator circuitry, and is useful while cascading two or more stages of the dot product engine. However, as presented in the next section, Config-B has a small non-zero current corresponding to zero input as opposed to Config. A that has zero current for zero input.

In order to sense the output current at the RBLs, we use a current to voltage converter. This can most simply be a resistor, as shown in Fig. \ref{fig:8t}. However, there are a few constraints. As the output current increases, the voltage drop across the output resistor increases, which in turn changes the desired current output. A change in the voltage on the RBL would also change the voltage across the transistors $M1$ and $M2$, thereby making their conductance a function of the voltage on the RBL. Thus, at higher currents corresponding to multiple rows of the memory array, the $I_{RBL}$ does not approximate the vector-matrix dot product, but deviates from the ideal output. This dependence of the RBL voltage on the current $I_{RBL}$ will be discussed in detail in the next section with possible solutions. 


\begin{figure*}[t]
		\centering
		\includegraphics[width=6.4in,keepaspectratio]{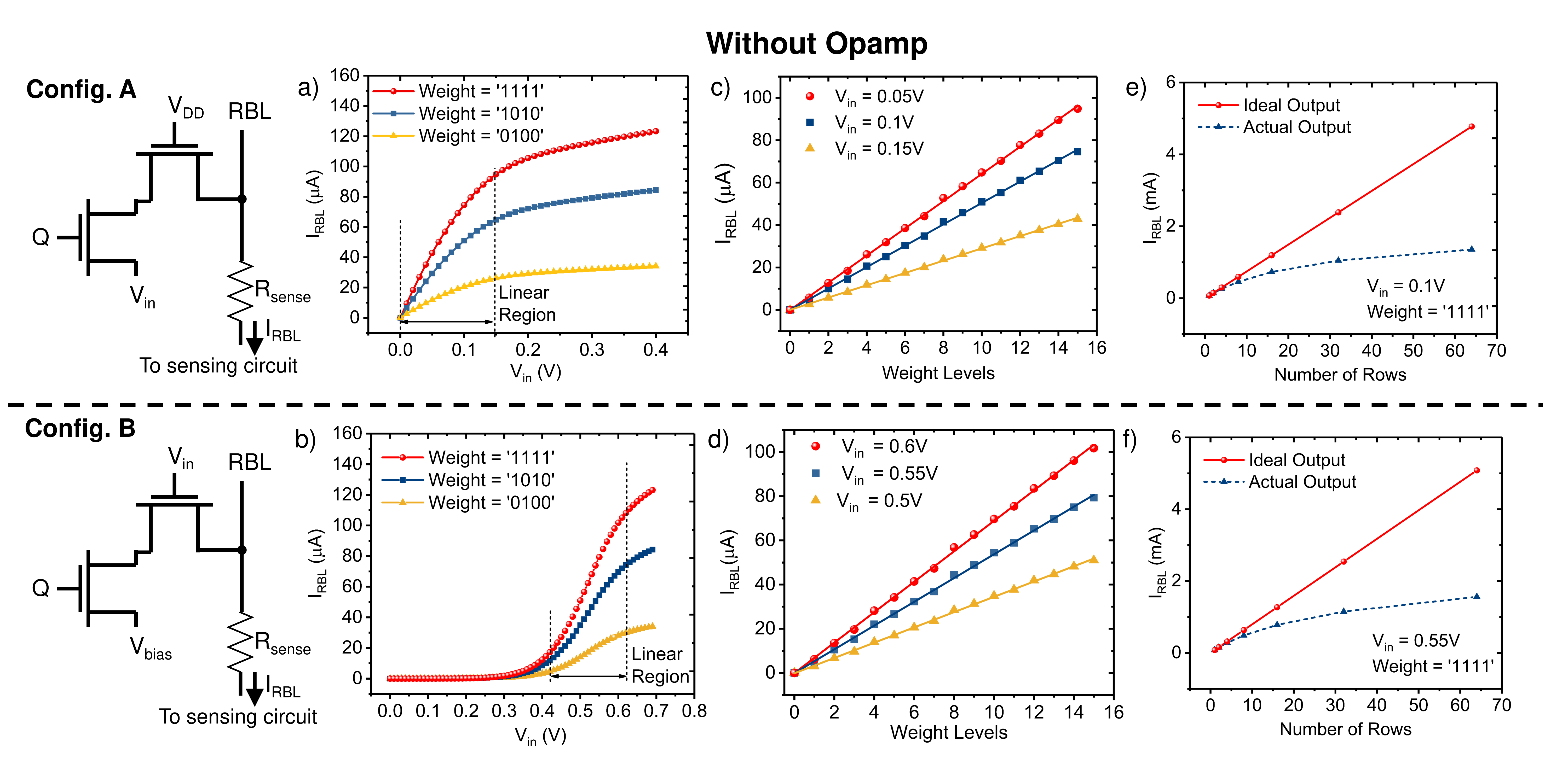}
		\caption{$I_{RBL}$ versus $V_{in}$ characteristics for (a) Config. A and (b) Config. B shows the linear region of operation for different weights. $I_{RBL}$ versus Weight levels for (c) Config. A and (d) Config. B shows desirable linear relationship at various voltages $V_{in}$. $I_{RBL}$ shows significant deviation from ideal output ($I_N = N\times I_1$ with increasing number of rows for both (e) Config. A and (f) Config. B, where $I_1$ is the current corresponding to one row and N is the number of rows. The analyses were done for $V_{DD} = 0.65V$} 
		\vspace{-4mm}
		\label{fig:compwoopamp}
\end{figure*}

\section{Results }

The operation of the proposed configurations (Config-A and Config-B) for implementing a multi-bit dot product engine was simulated using HSPICE on the 45nm PTM technology \cite{zhao2006new}. For the entire analysis, we have used a scaled down $V_{DD}$ of 0.65V for the SRAM cells. The main components of the dot-product engine implementation are the input voltages and conductances of the transistors for different states of the cells. A summary of the analysis for the two configurations is presented in Fig. \ref{fig:compwoopamp}. In Fig. \ref{fig:compwoopamp}, we have assumed a sensing resistance of 50-ohms connected to the RBL. Note, a small sense resistance is required to ensure that the voltage across the sensing resistance is not high enough to drastically alter the conductances of the connected transistors $M1$ and $M2$.

In Fig. \ref{fig:compwoopamp}(a)-(b) we plot the output current in RBL ($I_{RBL}$) as a function of the input voltage for three 4-bit weight combinations `1111', `1010' and `0100' for the two different configurations described in the previous section. The results presented are for a single 4-bit cell. To preserve the accuracy of a dot-product operation, it is necessary to operate the cell in the voltage ranges such that the current is a linear function of the applied voltage $v_i$. These voltage ranges are marked as linear region in Fig. \ref{fig:compwoopamp}(a)-(b). The slope of the linear section $I_{RBL}$ versus $V_{in}$ plot varies with weight, thus signifying a dot product operation. Further, at the left voltage extremity of the linear region, $I_{RBL}$ tends to zero irrespective of the weight, thus satisfying the constraint that the output current is zero for zero $V_{in}$. It is to be noted that the two configurations show significantly different characteristics due to the different point-of-application of input voltages.

Fig. \ref{fig:compwoopamp}(c)-(d) presents the dependence of the current $I_{RBL}$ on the 4-bit weight levels for Config-A at constant voltages $V_{in}$ = 0.05V, 0.1V, 0.15V and configuration B at $V_{in}$ = 0.5V, 0.55V, 0.6V, respectively. Different voltages were chosen so as to ensure the circuit operates in the linear region as depicted by Fig. \ref{fig:compwoopamp}(a)-(b). Desirably, $I_{RBL}$ shows a linear dependence on weight levels and tends to zero for weight = `0000'. The choice of any voltage in the linear regions of Fig \ref{fig:compwoopamp}(a)-(b) does not alter the linear dependence of the $I_{RBL}$  on weight levels. 

To expand the dot-product functionality to multiple rows, we performed an analysis for upto 64 rows in the SRAM array, driven by 64 input voltages. In the worst case condition, when the 4-bit weight stores `1111', maximum current flows through the RBLs, thereby increasing the voltage drop across the output resistance. Fig. \ref{fig:compwoopamp}(e)-(f) indicates that the total current $I_{RBL}$  deviates from its ideal value with increasing number of rows, in the worst case condition. The deviation in Fig. \ref{fig:compwoopamp}(e)-(f) is because we sense the output current with an equivalent sensing resistance ($R_{sense}$) and hence the final voltage on the bit-line ($V_{BL}$) is dependent on the current $I_{RBL}$. At the same time, $I_{RBL}$ is also dependent on $V_{BL}$ and as a result the effective conductance of the cell varies as $V_{BL}$ changes as a function of the number of rows. It was also observed that the deviation reduces with decreasing sensing resistance as expected. Another concern with respect to Fig. \ref{fig:compwoopamp} is the fact that the total summed up current reaches almost 6mA for 64 rows for the worst case condition (all the weights are `1111'). 
 \begin{figure*}[t]
		\centering
		\includegraphics[width=6.4in,keepaspectratio]{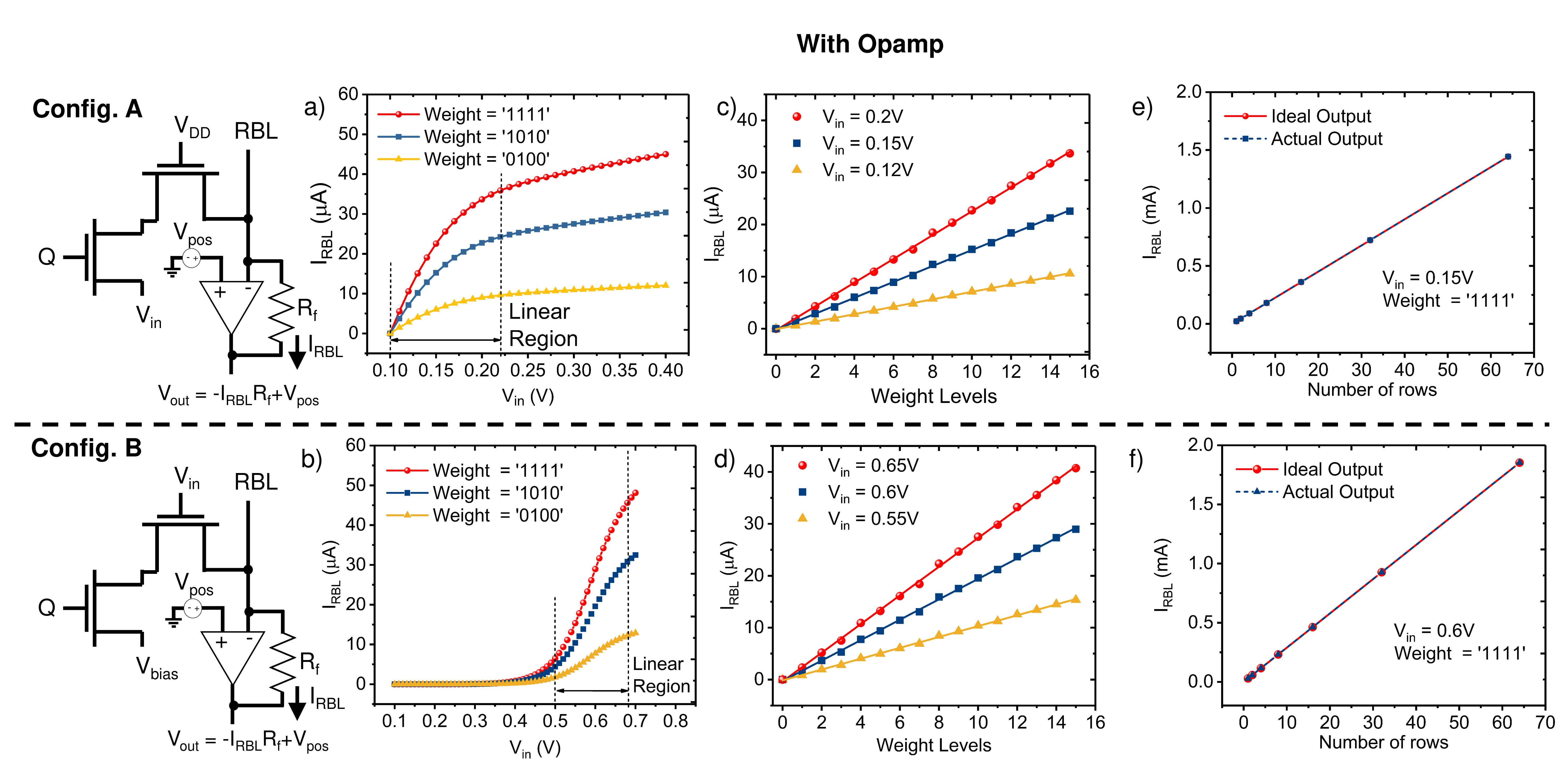}
		\caption{$I_{RBL}$ versus $V_{in}$ characteristics for (a) Config. A and (b) Config. B shows the linear region of operation for different weights. $I_{RBL}$ versus weight levels for (c) Config. A and (d) Config. B shows desirable linear relationship at various voltages $V_{in}$. $I_{RBL}$ shows almost zero deviation from ideal output ($I_N = N\times I_1$ with increasing number of rows for both (e) Config. A and (f) Config. B, where $I_1$ is the current corresponding to one row  and N is the number of rows. These analyses were done for $V_{DD} = 0.65V$} 
		\vspace{-4mm}
		\label{fig:compopamp}
\end{figure*}
 
There are several ways to circumvent the deviation from ideal behavior with increasing number of simultaneous row accesses and also reduce the maximum current flowing through the RBLs. One possibility is to use an operational amplifier (Opamp) at the end of each 4-bit column, where the negative differential input of the Opamp is fed by the bit-line corresponding to a particular column. Whereas, the positive input is supplemented by a combination of the Opamp offset voltage and any desired voltage required for suitable operation of the dot-product as shown in left hand side of Fig. \ref{fig:compopamp}. Opamp provides a means of sensing the summed up current at the RBL while maintaining a constant voltage at the RBL. Opams in the configuration as shown in Fig. \ref{fig:compopamp} have been traditionally used for sensing in memristive crossbars as in \cite{Li_2017}.

We performed the same analysis as previously described in Fig. \ref{fig:compwoopamp} for the two proposed configurations with the bit-line terminated by an Opamp. For our analysis, we have set $V_{pos} = 0.1V$ for the positive input of the Opamp and thus analysis is limited to input voltages above $V_{pos}$ to maintain the unidirectional current. Note, we have used an ideal Opamp for our simulations, where the voltage $V_{pos}$ can be accounted for both the non-ideal offset voltage of the Opamp and a combination of an externally supplied voltage. Fig. \ref{fig:compopamp}(a)-(b) shows the plot of $I_{RBL}$ versus input voltage $V_{in}$ for the two configurations. Similar behavior as in the case of Fig. \ref{fig:compwoopamp}(a)-(b) is observed even in the presence of the Opamp. However, note that the current ranges have decreased since RBL is now clamped at $V_{pos}$. Further, the dot-product operation is only valid for $V_{in}>V_{pos}$ and thus the acceptable input range is shifted in the presence of an Opamp. Fig. \ref{fig:compopamp}(c)-(d) shows the behavior of $I_{RBL}$ versus weight levels for the two configurations and desirably, linearity is preserved. 

Fig. \ref{fig:compopamp}(e)-(f) presents the current through the RBL as a function of the number of rows. As expected, 
due to the high input impedance of the Opamp, and  the clamping of $V_{BL}$ at a voltage $V_{pos}$ the deviation of the summed up current from the ideal value have been mitigated to a huge extent. Although, the current levels have reduced significantly as compared to the Fig. \ref{fig:compwoopamp}, the resultant current for 64 rows would still be higher than the electro-migration limit for the metal lines constituting the RBL \cite{posser2014analyzing}. One possible solution is to sequentially access a smaller section of the crossbar (say 16 or 8 rows at a time), convert the analog current into its digital counterpart each time and finally add all accumulated digital results. In addition use of high threshold transistors for the read port of the SRAM would also help to reduce the maximum current values. Further, the maximum current is obtained only when all the weights are `1111', which is usually not true due to the sparsity of matrices involved in various applications as in \cite{changpinyo2017power,han2015learning}.

\begin{figure}[t]
\centering
\includegraphics[width=0.3\textwidth]{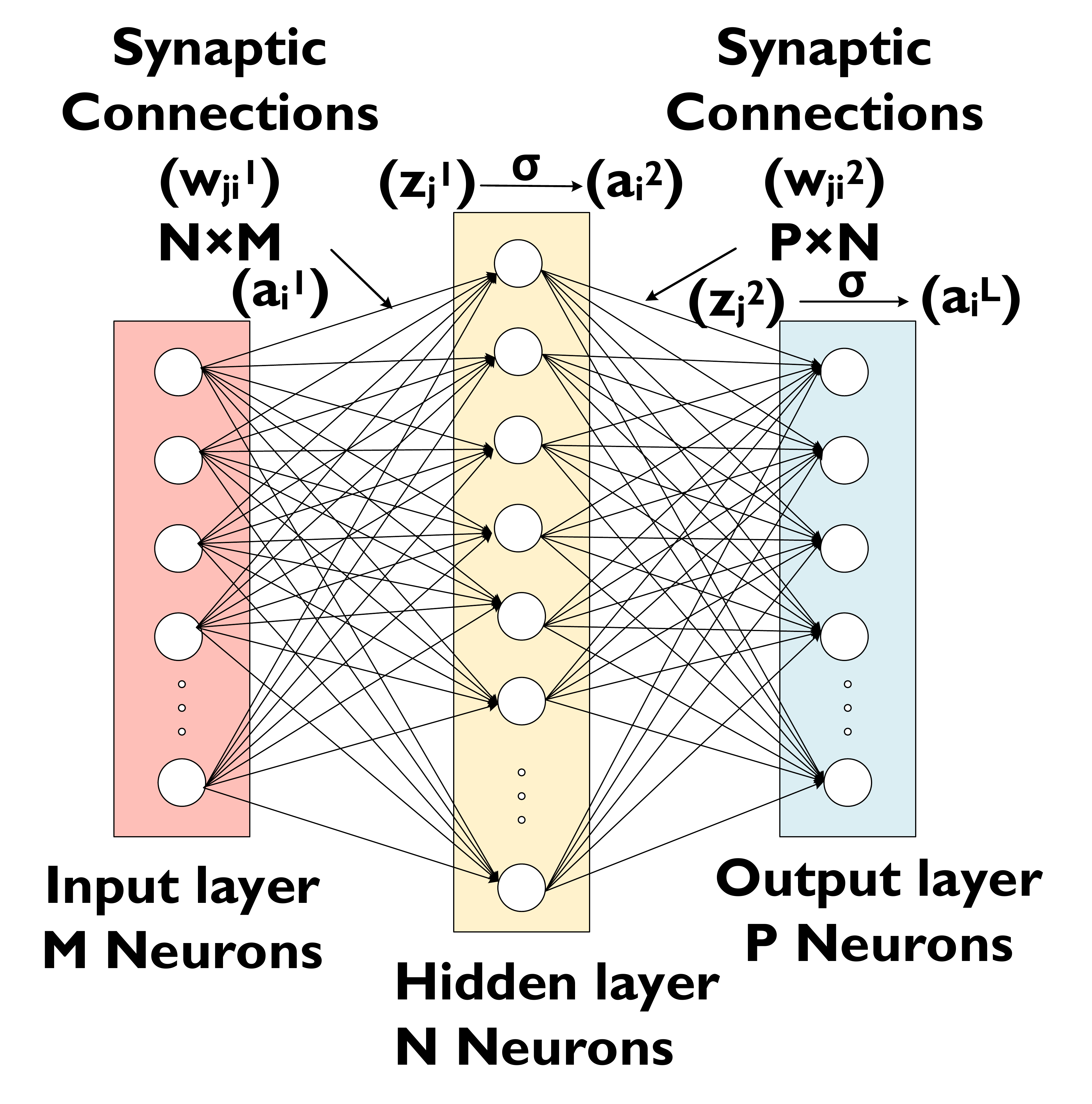}
\caption{Fully connected network topology consisting of 3 layers, the input layer, the hidden layer and the output layer \cite{chakraborty2017technology}. We have used M=784, N = 500 and P = 10.
}
\label{fig:fcn} 
\end{figure}

\par 
We also performed functional simulations using the proposed dot-product engine based on Config. A in a fully connected artificial neural network consisting of 3 layers as shown in Fig. \ref{fig:fcn}. The main motivation behind this analysis is to evaluate the impact of the non-linearity in the I-V characteristics on the inference accuracy of the neural network. We chose an input voltage range of 0.1-0.22V. As can be observed in Fig. \ref{fig:compopamp}(a), the I-V characteristics are not exactly linear within this range, as such a network level functional simulation is required to ascertain the impact of the non-linearity on classification accuracy. The network details are as follows. The hidden layer consisted of 500 neurons. The network was trained using the Backpropagation algorithm \cite{Rumelhart_1985} on the MNIST digit recognition dataset under ideal conditions using MATLAB \textsuperscript\textregistered Deep Learning
Toolbox\cite{palm2012prediction}. 

During inferencing, we incorporated the proposed 8T-SRAM based dot-product engine in the evaluation framework by discretizing and mapping the trained weights proportionally to the conductances of the 4-bit synaptic cell. The linear range of the voltage was chosen to be [0.1-0.22V] and normalized to a range of [0 1]. The dot-product operation was ensured by normalizing the I-V characteristics for all the weight levels such that current corresponding to the highest input voltage and highest weight level is $I_{max} = V_{max}\times G_{max}$. The activation function of the neuron was considered to be a behavioral $satlin$ function scaled according to the scaling factor of the weights to preserve the mathematical integrity of the network. To be noted, the normalization of current and input voltage simplifies the scaling of the neuron activation function. The accuracy of digit recognition task was calculated to be merely 0.11\% lower than the ideal case (98.27\%) thus indicating that the proposed dot-product engine can be seamlessly integrated into the neural network framework without significant loss in performance.
\par 
\begin{figure}[t]
\centering
\includegraphics[width=0.5\textwidth]{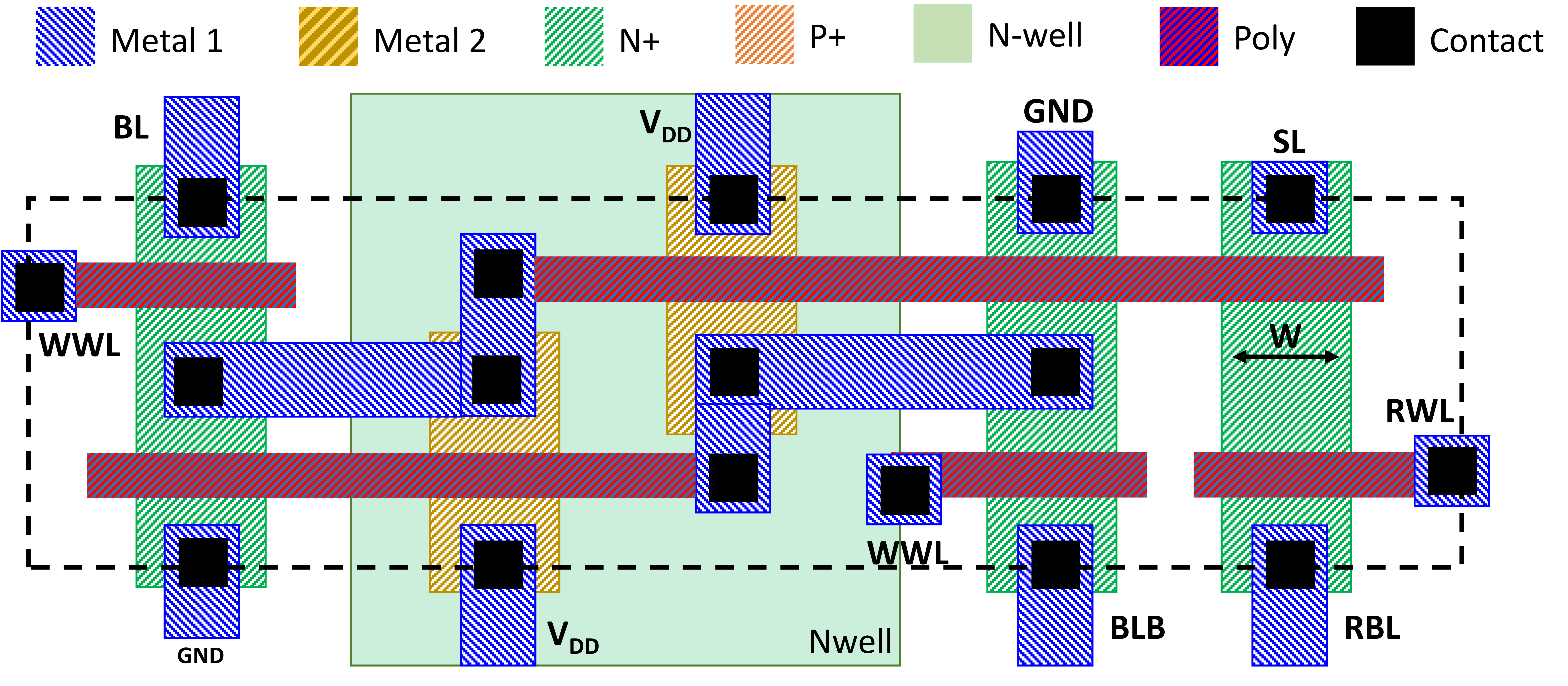}
\caption{Thin-cell layout for a standard 8T-SRAM bit-cell \cite{chang2005stable}.
}
\label{fig:layout_std} 
\end{figure}

\begin{figure*}[t]
\centering
\includegraphics[width=\textwidth]{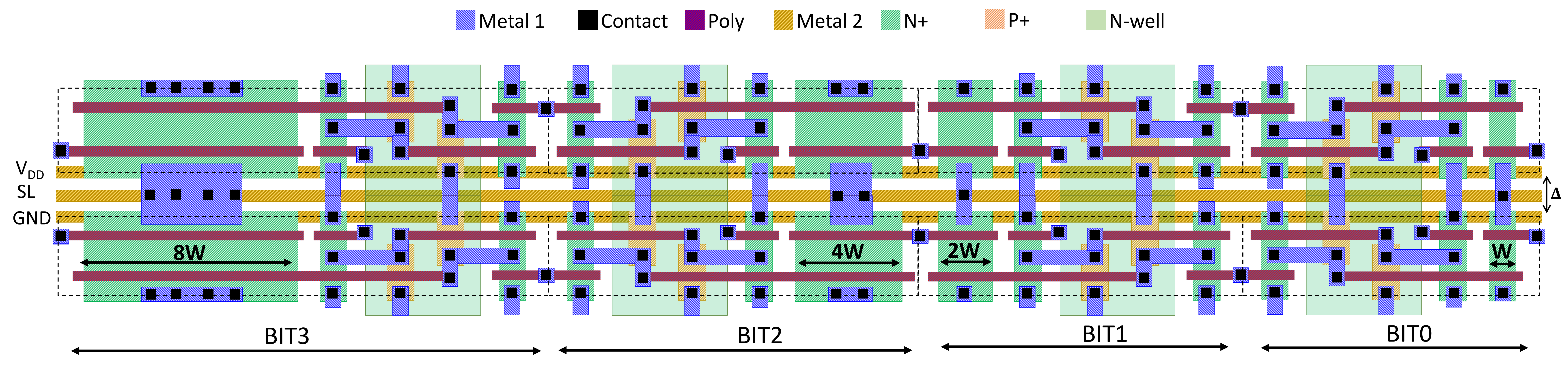}
\caption{Thin-cell layout for the proposed 8T-SRAM array with 4-bit precision weights. The width of read transistors of different bit positions are sized in the ratio 8:4:2:1. An additional metal line for SL is also required, which runs parallel to the power-lines. This incurs an area overhead of $\sim$29.4\% compared to the standard 8T-SRAM bit-cell.
}
\label{fig:layout} 
\end{figure*}

Further, it is to be noted that in many cases the inherent resilience of the applications that require dot product computations can be leveraged to circumvent some of the circuit level non-idealities. For example, for cases like training and inference of an artificial or a spiking neural network, various algorithmic resilience techniques can be applied where modeling circuit non-idealities and modifying the standard training algorithms\cite{chakraborty2017technology, Liu_2017} can help preserve the ideal accuracy of the classification task concerned.
Additionally, the proposed technique can either be used as a dedicated CMOS based dot product compute engine or as an on-demand dot product accelerator, wherein the 8T array acts as usual digital storage and can also be configured as a compute engine as and when required. It is also worth mentioning that the 8T cell has also been demonstrated in \cite{agrawal2017x} as a primitive for vector Boolean operations. This work, significantly augments the possible use cases for the 8T cells by adding analog-like dot product acceleration.

Due to different sizing of the read transistors and an additional metal line routing for SL, there is an area penalty of using the proposed configurations, compared to the standard 8T-SRAM bit-cell used for storage. Fig. \ref{fig:layout_std} shows the thin-cell layout for a standard 8T-SRAM bit-cell \cite{chang2005stable}. Note that the rightmost diffusion with width (W) constitute the read transistors ($M1$ and $M2$). To implement the 4-bit precision dot-product, we size the width of read transistors in the ratio $8:4:2:1$, as described earlier. Thus, the width of the rightmost diffusion is increased to 8W, 4W, and 2W, increasing the bitcell length (horizontal dimension) by $\sim$39.6\%, 17.1\%, and 5.7\% for bits 3, 2 and 1, respectively, compared to the standard minimum sized 8T bit-cell with diffusion width W. Moreover, to incorporate an extra metal line (SL), that runs parallel to $V_{DD}$ and ground lines, the cell width (vertical dimension) increases by $\sim$12.5\%. The resulting layout of first four columns for two consecutive rows in the proposed array is shown in Fig. \ref{fig:layout}. The overall area overhead for the whole SRAM array with 4-bit weight precision, amounts to $\sim$29.4\% compared to the standard 8T SRAM array. Note, this low area overhead results from the fact that both the read transistors $M1$ and $M2$ share a common diffusion layer and hence an increase in transistor width can be easily accomplished by having a longer diffusion, without worrying about spacing between metal or poly layers. Additionally, instead of progressively sizing the read transistors one could also use multi-V$_T$ design wherein the LSBs consist of hight V$_T$ read transistors and the MSBs consist of nominal ( or low V$_T$ read transistors). The use of multi-V$_T$ design can significantly reduce the reported area overhead. As such, the reported area overhead is close to the worst case impact on the bit-cell area without resorting to additional circuit tricks like multi-V$_T$ design. 
\par

\section{Variation Analysis}
To ascertain the robustness of the presented dot product computations, in this section, we analyze the effects of non-idealities on the output current. The non-idealities considered are SL and BL line-resistances and transistor threshold voltage variations. 

\subsection{Effect of Line-Resistances}
\begin{figure*}[t]
	\centering
	\includegraphics[width=\textwidth,keepaspectratio]{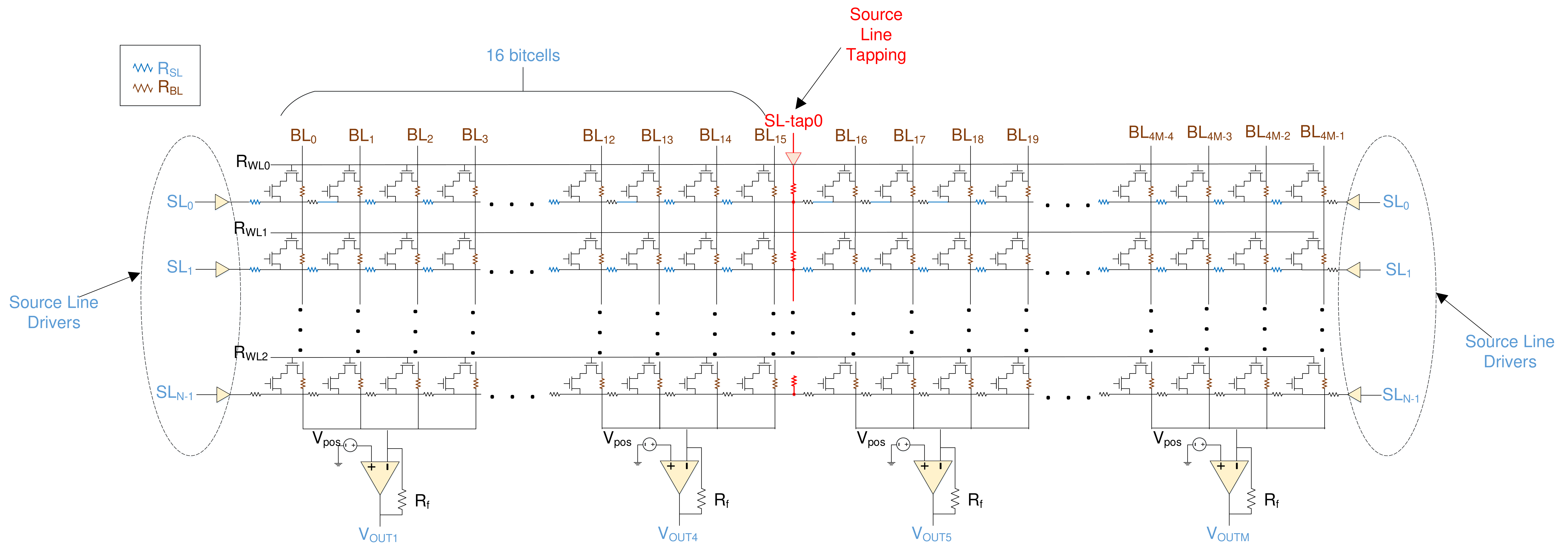}
	\caption{Bitcell organization of Config. B variants showing SL driven from both ends and tapping of SL every 16 bitcells. The line resistances in the source line (SL) and the bit-line (BL) are shown. }
	\label{fig:lineresfig}
\end{figure*}

Both the SL and BL line-resistances add parasitic voltage drops along the rows and the columns. Moreover, to complicate the analysis, the error in the output current would be a function of both the spatial dependence due to distributed line-resistances and data-dependence as a function of the stored weights in the memory array. We, therefore, resort to worst case analysis. The worst case arises when all the weights and all the inputs are at the highest value. This scenario results in maximum current flow through the BLs and SLs and hence has maximum impact of parasitic line-resistances. To analyze the impact, we consider a line resistance of 1.3 ohms/$\mu m$ \cite{thoziyoor2008cacti}. Based on the layout, the average line resistance between each bit-cell was found to be 1.25 ohms in the bit-line direction and 2.5 ohms in the SL direction. We explore both the configurations (Config. A and Config. B) to analyze the impact of the line-resistances and ways to compensate for the voltage degradation along the metal lines. In addition, for Config. B, we explore two variants to minimize the effect of line resistances. Note, in Config. B the inputs are connected to the word-lines i.e. to the gate of the transistors. As such, the inputs drive capacitive load and there is no voltage degradation due to line-resistances. On the other hand, the bias voltage is connected to the SL, which would degrade due to line-resistances and induce error in the final output current flowing through each column. To minimize this error, the two variants of Config. B presented in the manuscript are:
\begin{itemize}
	\item Config. B with the bias voltage driving the SL from both the ends (i.e. from the extreme right and extreme left ends, as shown in Fig. \ref{fig:lineresfig}).
	\item Config. B with the SL tapped every 16 bits with regenerated values of the bias voltage in the horizontal direction, as depicted in Fig. \ref{fig:lineresfig}. 
\end{itemize}
 
\begin{figure}[t]
	\centering
	\includegraphics[width=2.5in,keepaspectratio]{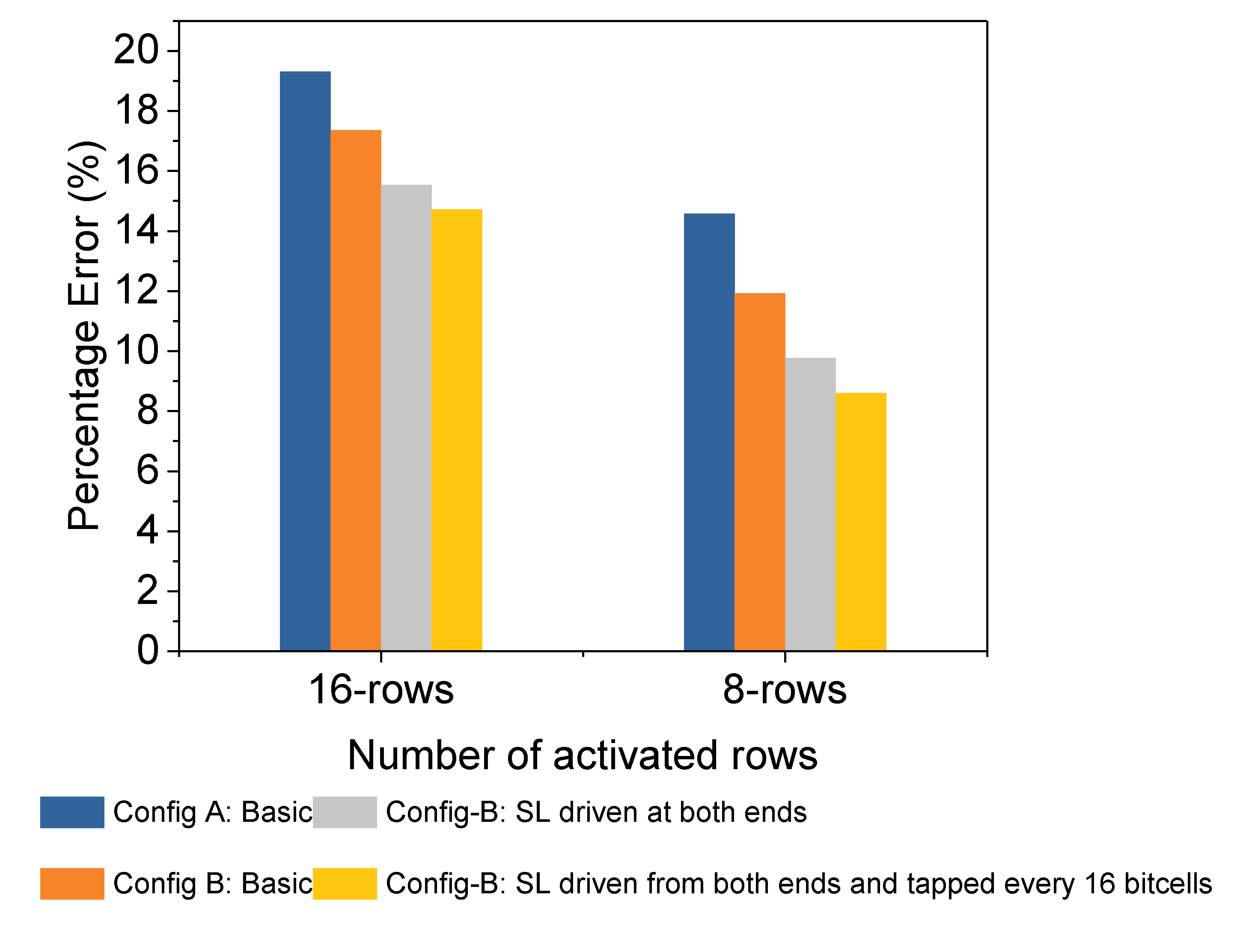}
	\caption{ Percentage error in output current for worst case combination (highest input values and all weights = `1111').  The left set of bar graphs represent the error for various combinations assuming 16 rows are activated simultaneous for the dot product computation, while the right set of bar graphs correspond to simultaneous activation of 8 rows.}
	\label{fig:lineerrordiffcomb1}
\end{figure}

Fig. \ref{fig:lineerrordiffcomb1} shows the worst case impact (when all inputs are at the highest value and all the weights are `1111') of the line-resistances in terms of percentage error in the output current (Note, this is error with respect to the current values, it should not be confused with the error corresponding to the classification accuracy) for the various configurations for simultaneous activation of 16 and 8 rows, respectively. As observed, Config. A has a higher error than all the variants of Config. B. Note, tapping is infeasible in case of Config. A because in Config A, the input voltages are connected to the SL. Tapping in Config. A would therefore require regeneration of input voltage along the horizontal direction, making it infeasible. In contrast, in case of Config. B, SL is supplied by a global bias voltage and hence, is easy to regenerate. We have assumed an array size of 64x128 (64 rows and 128 columns). Further, for our analysis we assume that the `farthest' 16 rows are simultaneously activated. SL and BL distributed resistances were included for all the activated rows, while the unactivated rows were modeled by an equivalent lumped resistance.

  \begin{figure}[t]
	\centering
	\includegraphics[width=3.5in,keepaspectratio]{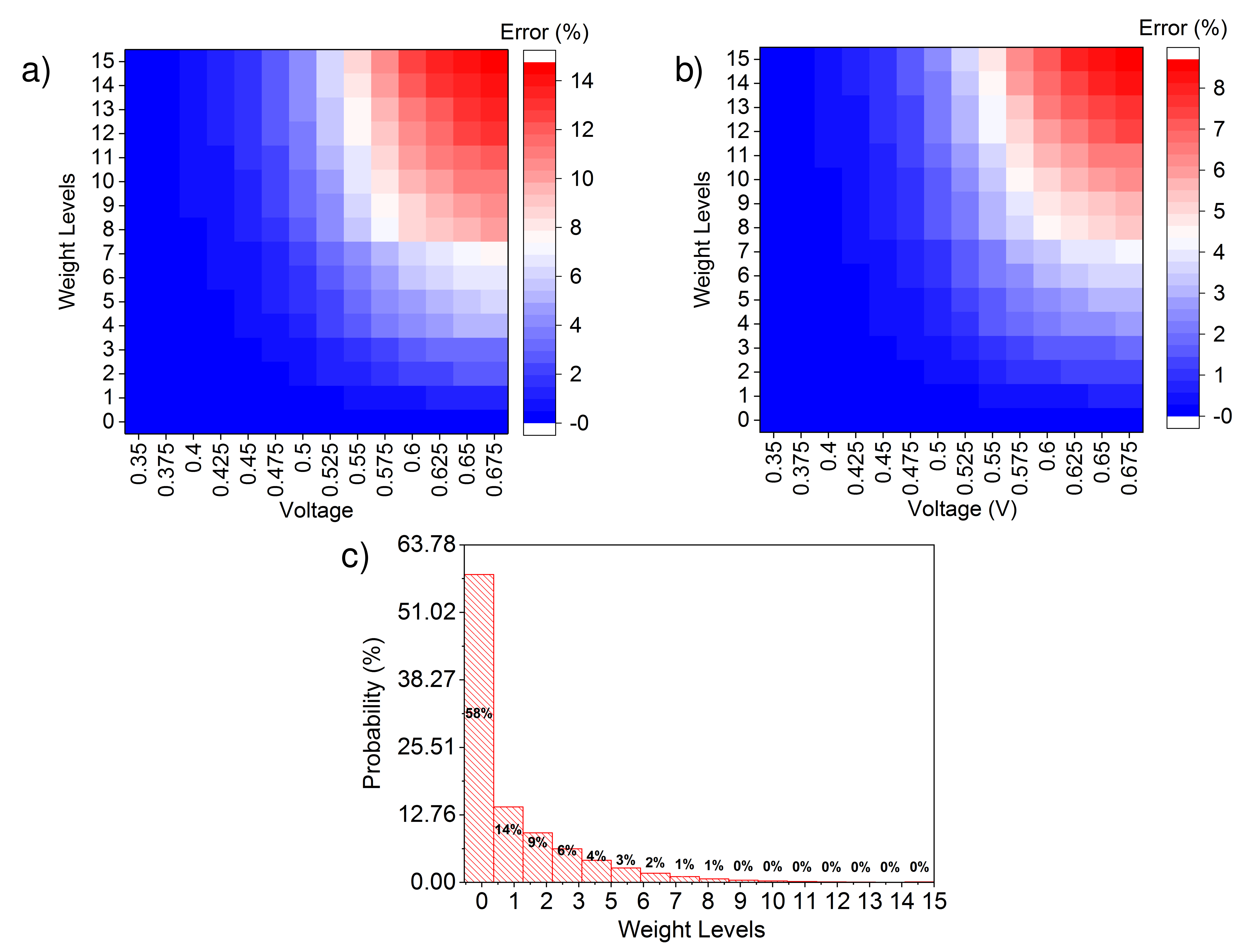}
	\caption{(a) and (b) shows the percentage error map arising due to line resistance for different weight levels ranging from `0000' to `1111' and input voltages ranging from 0.35V to 0.675V for 16 and 8 activated rows. For e.g., the data point corresponding to V = 0.35V and weight level = `0000' means the test case where all the 4-bit weight elements in the memory array are considered to be at weight `0000' and the input voltages to all rows are 0.35V. The percentage error decreases with decreasing weight and input value. (c) Probability of occurrence of weight levels in a trained neural network on MNIST dataset shows lowest weight levels have the highest frequency, thus indicating low impact due to line resistance.}
	\label{fig:lineres1}
\end{figure}

For rest of the analysis, we choose Config. B with tapping every 16 bits. We now analyze the error percentage across all voltages and weight combinations to understand the impact of the degradation in light of applications discussed in the manuscript. Fig. \ref{fig:lineres1} (a) and (b) shows the 2D error-map due to line resistances for various combinations of input voltages and weights for 16 and 8 activated rows, respectively. Note, for each input voltage $-$ weight combination all rows are supplied with the same input voltage and all weights in the array are same. In addition, Fig. \ref{fig:lineres1} (c) shows the weight level distribution of a neural network layer trained on the MNIST dataset. As observed from Fig. \ref{fig:lineres1} (a) and (b), the error above 6\% for 16 rows and above 4\% for 8 rows is concentrated to the top 25\% of the map corresponding to the highest weights and inputs. However, from Fig. \ref{fig:lineres1} (c), we observe that for relevant applications such as neural network the trained weights are mostly concentrated to the weight level 1-6 where the error is close to 0-5\% for 16 rows and 0-3.4\% for 8 rows. From this analysis, we can conclude that using the circuit techniques presented \textit{i.e.} driving SL from both the sides and tapping every 16 columns and also leveraging the weight distribution for a trained neural network, the effect of line resistances for simultaneous activation of 8 and 16 rows can be substantially mitigated. For example, for Config. B with taps every 16 columns with SL being driven from both the ends, the worst-case error is contained within 9\%. Further, it was observed that the error improves rapidly when the input voltages or the programmed weights are less than their maximum possible values. 

\subsection{$V_T$ Variations}
The variations in transistors can result in error in the dot-product operations. To analyze this, we perform 1000 Monte-Carlo simulations to assess the variation of the output current for various combinations of input voltages and weights. We considered 30 mV $\sigma$ variation of threshold voltage ($V_T$)for the minimum sized transistor and scaled the variation with width as $\sigma_L = \sigma_{min}\sqrt{W_{min}L_{min}/WL}$. Note, for random variations it is customary to include various sources of variations into effective variation in the transistor threshold voltage \cite{kuhn2011process}.  We ran 1000 Monte-Carlo simulations for each voltage value ranging from 0.35V to 0.675V in steps of 0.025V and each weight level ranging from 0 to 15 and obtained the standard deviation in output current for each case. This captures the impact of Vth variations for a considerable precision of gate voltages. We calculated the standard deviation about the mean current for the entire range of output current from the cases described above for 16 activated rows of the memory array. The minimum current on the x-axis in Fig. \ref{fig:sd1} arises when the input voltages and (or) the stored weights in the memory array are zero. The next higher level of current is obtained when either the weight or the input voltage is incremented. It is worth noting that Fig. \ref{fig:sd1} corresponds to 16 activated rows in an array of 64 rows and 128 columns. Further, for the analysis in Fig. \ref{fig:sd1}, we have neglected the effect of line-resistances for the following reasons - 1) adding line-resistances makes the standard deviation in Fig. \ref{fig:sd1} a function of not only the random $V_T$ variation and weights, but also makes the deviation in current spatially dependent. This leads to a non-trivial analysis problem that can quickly become intractable. 2) As shown in the previous sub-section, even the worst case error due to line-resistances was well within acceptable limits. 
Fig. \ref{fig:sd1} shows that the standard deviation is higher for a higher value of output current. We fit a representative standard deviation for each current value using a polynomial fit as shown by the fitted line in Fig. \ref{fig:sd1}. 
\begin{figure}[t]
	\centering
	\includegraphics[width=3in, keepaspectratio]{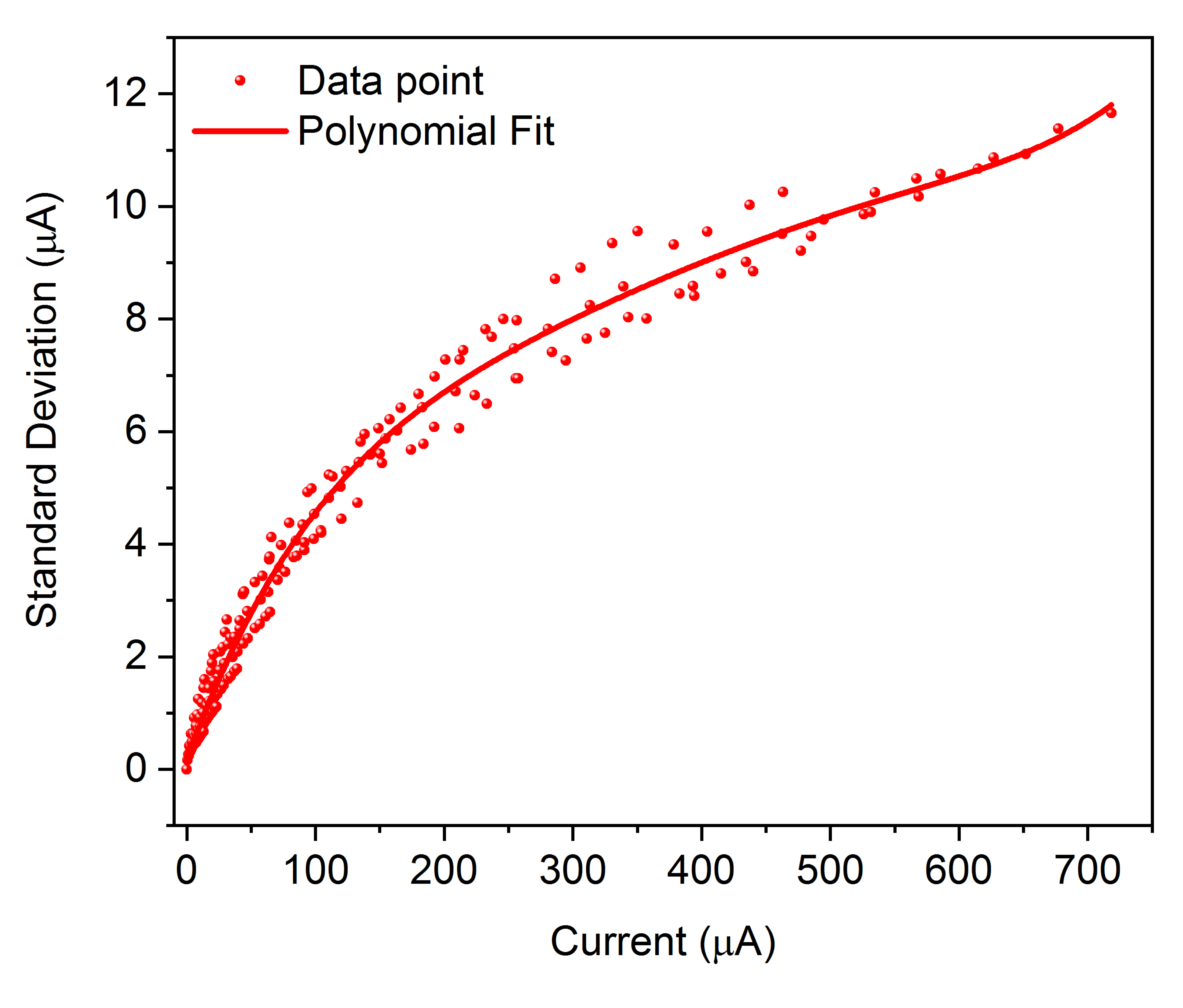}
	\caption{Standard deviation of current due to variations in Vt of the transistors of the bitcells with increasing current for 1000 Monte Carlo simulations. A single data point shown here refers to the standard deviation in output current when 16 rows are activated and input voltages to all rows are $V_{in}$ and weights of all elements are $w$. For different data points, we consider $V_{in}$ values ranging from 0.35V to 0.675V in steps of 0.025V and weight levels ranging from 1 to 16 to capture the impact of $V_T$ variations across the input parameter space.}
	\label{fig:sd1}
\end{figure}
In the functional simulations to evaluate the classification accuracy with such $V_T$ variations, we calculated the output current from every 16 rows and replaced it with a random value from a Gaussian distribution with the corresponding standard deviation of the particular output current. The final classification accuracy was 0.01\% lower than the case without random variations. This error resilience is mainly due to the robustness of the neural networks and the fact that as in Fig. 10 most of the weights are concentrated in lower levels that have lesser standard deviation.

\section{Discussions}
We would like to emphasize the fact that the present proposal aims at providing a means to enable \textit{in-situ} dot-product computations in standard 8T SRAM cells by exploiting the isolated read-port. We believe a wide-range of applications can be accelerated using the present proposal. As such, the presented dot product engine should not be seen only in context of machine learning and neural network applications. In general, any application that can benefit from approximate vector addition and multiplication can be a possible use case for the presented proposal. This wide spectrum of possible use cases implies that the exact details of the required peripheral circuits and its complexity would depend heavily on the target application. For example, error resilient applications like neural networks can rely on low cost peripherals whereas more traditional dot-product computations as in image processing might require sophisticated circuitry. Moreover, one could think of hybrid significance driven peripheral design such that the less significant computations are associated with low overhead peripherals while more significant operations are enabled by high accuracy circuits or a full digital computation without resorting to dot product acceleration. The target application would also dictate the constrains on OPAMP specification and the required precision of the resistance Rf shown in Fig. 8. In addition, the choice of Config. A versus Config. B would also depend on the target application. For example, Config. A shows better linearity as opposed to Config. B. However, the input voltages in Config. A drive a resistive load requiring complex driving circuits as opposed to Config. B which has capacitive load. The authors would also like to point that a detailed analysis of the appropriate peripherals and the associated architecture for each individual use case requires a case by case analysis and is not the focus of the present manuscript. The current manuscript is more of a generic proposal and a study of the effect of intrinsic non-idealities, for example, the non-linearity, the line-resistances and the transistor threshold voltage variations with respect to the present dot product engine.

\begin{figure}[t]
	\centering
	\includegraphics[width=1.9in, keepaspectratio]{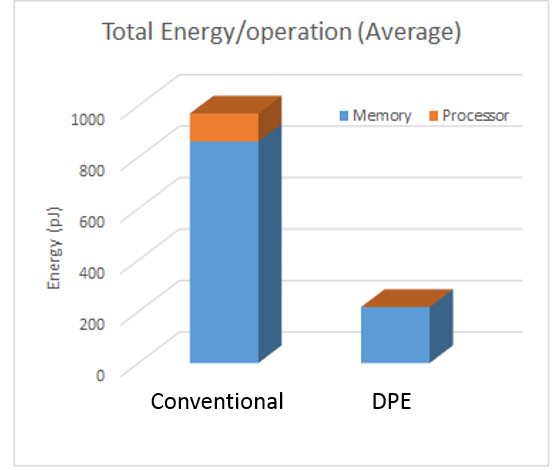}
	\caption{ Average Energy comparison between conventional digitial sequential implementation and proposed Dot-product Engine (DPE). The energy is reported for 16x16 dot product computations wherein 16 rows are simultaneously activated and each row consists of 16 4-bit words.}
	\label{fig:sd1}
\end{figure} 

We would now present the estimates for energy consumptions by performing 16x16 dot product operation with and without the proposed dot product engine. It is worth mentioning that the overall dot-product engine consists of DACs to generate analog inputs fed to the 8T-SRAM crossbar array, along with ADCs to detect the analog outputs and converting them back to digital bits. A cache memory of size 256KB with a basic sub-array size of 64$\times$128 bits was modeled using CACTI \cite{muralimanohar2009cacti} simulator. The energy consumption and latency of the peripheral circuitry (ADCs and DACs) was appropriately incorporated in the CACTI model, referring to \cite{liu201010b}. We assume a 16$\times$16 crossbar operation (\textit{i.e.} activating 16 rows at a time with each row containing 16 four bit words) at any given time, thus requiring 16 ADCs in the peripheral circuitry, per sub-array. The conversion time for the ADC operation was assumed to be 10ns and the energy estimates for the ADCs were adopted from \cite{liu201010b}. This framework was used to evaluate the total energy consumption and latency of the proposal for a test vector of 16$\times$16 dot-product, compared to the pure digital approach wherein the dot product was computed by sequential memory access and multiply as well as add operations were performed in dedicated adders and multipliers synthesized separately.

Fig. 12 shows the energy for performing a 16$\times$16 dot-product with the proposed DPE and the conventional digital approach. 
This energy overhead stems from the fact that in digital approach, row-by-row access to the data from memory, followed by MAC operations are performed sequentially to compute the same 16$\times$16 matrix-vector dot-product, which the proposed DPE can do in a single instruction.
Also, it was noted that the total energy consumption of the dot-product engine had a dominant contribution from the peripheral circuitry. Nevertheless, in general, the energy and latency overheads associated with respect to DACs and ADCs in similar dot product engines based on memristors have been extensively studied and can be found in works like \cite{shafiee2016isaac,chi2016prime}. 

\section{Conclusion}

In the quest for novel in-memory techniques for beyond von-Neumann computing, we have presented the 8T-SRAM as a vector-matrix dot-product compute engine.
Specifically, we have shown two different configurations with respect to 8T SRAM cell for enabling analog-like multi-bit dot product computations. We also highlight the trade-offs presented by each of the proposed configurations. The usual 8T SRAM bit cell circuit remains unaltered and as such the 8T cell can still be used for the normal digital memory read and write operations. The proposed scheme can either be used as a dedicated dot product compute engine or as an on-demand compute accelerator. The presented work augments the applicability of 8T cells as a compute accelerator in the view that dot products find wide applicability in multiple data intensive application and algorithms including efficient hardware implementations for machine learning and artificial intelligence.

\section*{Acknowledgement}
The research was funded in part by C-BRIC, one of six centers in JUMP, a Semiconductor Research Corporation (SRC) program sponsored by DARPA,  the National Science Foundation, Intel Corporation and Vannevar Bush Faculty Fellowship.

\bibliographystyle{IEEEtran}
\bibliography{bare_jrnl}

\end{document}